# Effect of the accuracy of interatomic force constants on the prediction of lattice thermal conductivity

Han Xie,[1] Xiaokun Gu,[2] and Hua Bao[1*]

[1]*University of Michigan-Shanghai Jiao Tong University Joint Institute,*

*Shanghai Jiao Tong University, Shanghai 200240, China*

[2]*Institute of Engineering Thermophysics, School of Mechanical Engineering,*

*Shanghai Jiao Tong University, Shanghai 200240, China*

**ABSTRACT**

Solving Peierls-Boltzmann transport equation with interatomic force constants (IFCs) from first-principles calculations has been a widely used method for predicting lattice thermal conductivity of three-dimensional materials. With the increasing research interests in two-dimensional materials, this method is directly applied to them but different works show quite different results. In this work, classical potential was used to investigate the effect of the accuracy of IFCs on the predicted thermal conductivity. Inaccuracies were introduced to the third-order IFCs by generating errors in the input forces. When the force error lies in the typical value of first-principles calculations, the calculated thermal conductivity would be quite different from the benchmark result. It is found that imposing translational invariance conditions cannot always guarantee a better thermal conductivity result. It is also shown that Grüneisen parameters cannot be used as a necessary and sufficient criterion for the accuracy of third-order IFCs in the aspect of predicting thermal conductivity.

* Corresponding author: hua.bao@sjtu.edu.cn

## 1. INTRODUCTION

In crystalline semiconductors and insulators, phonons (*i.e.* lattice vibrations) are the major heat carriers, so the thermal conductivity can be calculated with the knowledge of phonon properties. Recently there are growing interests to predict the lattice thermal conductivity by solving Peierls-Boltzmann transport equation (PBTE), either under single mode relaxation time approximation (SMRTA) or with full iterative solution.[1,2] With interatomic force constants (IFCs) extracted from first-principles calculations as the input, this method has been widely used to calculate the thermal conductivity of three-dimensional (3D) materials.[3–26] The good agreement between these calculated values and measured experimental data proves its accuracy and reliability.[3–21]

The discovery of ultrahigh thermal conductivity of graphene[27] has stimulated a growing research interest in the thermal conductivity of two-dimensional (2D) materials. The first-principles PBTE method for predicting thermal conductivity has been directly applied to 2D materials like graphene,[28–31] silicene,[32–35] phosphorene,[36–40] $MoS_2$,[41–43] borophene,[44] *etc.* However, previous calculations show quite different results in different works. For example, the predicted thermal conductivity for black phosphorene from SMRTA[36–40] at room temperature along armchair direction varies a lot from 5.46 to 33 W/mK, and the result along zigzag direction spans a rather large range between 15.33 and 83.5 W/mK. Similar discrepancy is also observed in the case of $MoS_2$, whose thermal conductivity at room temperature was predicted from SMRTA to be 83 W/mK by Li *et al.*[41] as well as Gu and Yang.[42] In comparison, Yan *et al.*[43] got a result of 35.5 W/mK with the same method. The lack of consistency in the predicted thermal conductivity may arise from using iterative method instead of SMRTA,[30,31] imposing translational invariance conditions to third-order IFCs,[15,35] different exchange-correlation functionals used in first-principles calculations,[45] enforcing a quadratic branch in the dispersion of 2D materials,[44] *etc*. In materials where resistive Umklapp scattering processes dominate in the phonon-phonon scattering, both SMRTA and iterative method should give similar results.[1,2] For graphene, iterative method yields a much larger thermal conductivity value than SMRTA because the momentum-conserving normal scattering processes also play an important role in phonon transport,[30,31] which has been explained by the hydrodynamic phonon transport at room temperature.[46,47] Regarding the translational invariance conditions (*i.e.* acoustic sum rules), Lindsay *et al.*[15] demonstrated that imposing it to third-order IFCs would play an important role in determining thermal conductivity. Our previous results for silicene also showed its importance.[32,35] Translational invariance conditions will affect the calculated phonon scattering rate, especially for long-wavelength phonons near the Brillouin zone center. Concerning the exchange-correlation functionals, Jain and McGaughey[45] studied their effects on predicting the thermal conductivity of crystalline silicon from first-principles calculations. Different exchange-correlation functionals could lead to an under-prediction or over-prediction within 20% of the experimental value. Recently, Carrete *et al.*[44] showed that the second-order IFCs directly extracted from first-principles calculations might yield problematic phonon dispersion curve for 2D materials. The dispersion curve of unstrained 2D materials can be demonstrated to contain a quadratic branch near Brillouin zone center[48] but the second-order IFCs directly extracted from first-principles calculations would yield linear ones. By generating physically sound IFCs, they showed that the thermal conductivity result could be quite different from the value calculated with the raw IFCs from first-principles calculations.[44]

Previous discussions explained part of the inconsistency in the predicted thermal conductivity of 2D materials from first-principles calculations. However, with all the above mentioned reasons considered, based on our own testing (unpublished), discrepancy in predicted thermal conductivity could still exist for some 2D materials. As will be shown later, the accuracy of third-order IFCs also plays an important role on the predicted thermal conductivity. For first-principles calculations, the accuracy might be affected by energy cutoff, *k*-point grid, reciprocal space projection technique, aliasing errors, discretization errors, *etc*.[49] For example, the attainable fractional precision[49] in forces using VASP is $10^{-4}$. Currently, the most widely used method to extract interatomic force constants generally takes force as the input parameter, and uses finite-difference method to calculate the IFCs. The raw IFCs from first-principles calculations often do not satisfy the translational invariance conditions, so these conditions are artificially imposed by adding small compensation to each term. All these processes will induce additional uncertainty to the IFCs, especially the third-order IFCs.

In this work, we will discuss how the accuracy of IFCs could affect the predicted lattice thermal conductivity values. Due to the large uncertainty in first-principles calculations, we used classical potential to do such an investigation. Classical potential has the advantage that it has an explicit analytical form, so that the error only comes from numerical computation and can be reduced to a negligible amount to get an "accurate" thermal conductivity result as benchmark. Based on the benchmark case from classical potential, inaccuracies are artificially introduced to third-order IFCs and effects of these inaccuracies on the predicted thermal conductivity are investigated. SMRTA is used to calculate thermal conductivity due to the computational cost consideration and also because phonon relaxation time is well defined with this approach. Previously, Grüneisen parameters have been used as a simple test for the accuracy of the third-order IFCs.[28] The applicability of this criterion is also examined. Our result will shed some light on predicting thermal conductivity from first-principles calculations. In what follows, we describe the simulation methods and details in Sec. 2. Simulation results for silicon, graphene, and silicene are shown in Sec. 3. Our conclusions are summarized and discussed in Sec. 4.

## 2. SIMULATION METHODS AND DETAILS

### 2.1 Single mode relaxation time approximation method

For a periodic crystal structure under equilibrium state, the potential energy can be expanded as the Taylor series[50,51]

$$U = U_0 + \frac{1}{2!}\sum_{ij,\alpha\beta}\Phi_{ij}^{\alpha\beta}u_i^\alpha u_j^\beta + \frac{1}{3!}\sum_{ijk,\alpha\beta\gamma}\Psi_{ijk}^{\alpha\beta\gamma}u_i^\alpha u_j^\beta u_k^\gamma + \ldots, \tag{1}$$

where $U_0$ is the equilibrium potential energy. $u_i^\alpha$, $u_j^\beta$, and $u_k^\gamma$ are the displacements of $i$-th atom in $\alpha$ direction, $j$-th atom in $\beta$ direction, and $k$-th atom in $\gamma$ direction, respectively. $\Phi_{ij}^{\alpha\beta}$ is the second-order IFC and $\Psi_{ijk}^{\alpha\beta\gamma}$ is the third-order IFC. Even higher-order IFCs are neglected in this equation. Physically correct second-order and third-order IFCs have to satisfy the point/space group symmetry relations, translational invariance conditions, and rotational invariance conditions, which are shown in the Appendix. Comparison of Grüneisen parameters from second-order and third-order IFCs has been used as a simple test for the accuracy of third-order IFCs. The calculation of Grüneisen parameters and the definition of relative difference are also shown in the Appendix. The force acting on each atom is $\boldsymbol{F}_i = -\nabla_i U$ and for a structure under equilibrium state, $\boldsymbol{F}_i = \boldsymbol{0}$. With the IFCs as the input, the thermal conductivity of semiconducting or insulating materials can be calculated from SMRTA with the following equation

$$\kappa_l^{\alpha\beta} = \sum_\lambda c_{ph,\lambda} v_\lambda^\alpha v_\lambda^\beta \tau_\lambda, \tag{2}$$

where $\lambda$ denotes different phonon modes that can be distinguished by wave vector $\boldsymbol{q}$ and phonon branch $\nu$. $c_{ph,\lambda}$ is the volumetric phonon specific heat. $v_\lambda^\alpha$ and $v_\lambda^\beta$ are the phonon group velocity in $\alpha$ and $\beta$ directions, respectively. $c_{ph,\lambda}$ and $\boldsymbol{v}_\lambda$ can be calculated with the second-order IFCs as the input.[52] $\tau_\lambda$ is the phonon relaxation time, *i.e.* the inverse of phonon scattering rate. In our calculation of $\tau_\lambda$, only phonon-phonon scattering is considered. More specifically, only three-phonon scattering is considered. $\tau_\lambda$ can be calculated with both second-order and third-order IFCs as the input. For more details about the method we refer the reader to Refs. [52–54].

**2.2 Simulation details**

The second-order and third-order IFCs as the input for SMRTA method were calculated from classical potential. Careful tests were carried out to reduce the numerical error. GULP package[55] was first used to optimize the primitive unit cell of silicon, graphene, or silicene. After that, a supercell was constructed and the lattice constant was re-optimized with LAMMPS package.[56] In this re-optimization process, the bisection method was used to find the lattice constant corresponding to the lowest energy state. Forces acting on each atom were computed from LAMMPS through the analytical derivatives of the potential function,[57] which would be free of truncation error coming from numerical differentiation. As the input for calculating IFCs, forces were output with sixteen significant digits to retain accuracy. In order to reduce the truncation error, fourth order accuracy method was used instead of central difference method (see the Appendix for the details) to compute the second-order and third-order IFCs with our own in-house code and revised THIRDORDER.PY,[54] respectively. These third-order IFCs were not modified by adding small compensation because they already satisfied translational invariance conditions to a reasonable extent. Point/Space group symmetry conditions were enforced and utilized to reduce computational cost. This result was then used to obtain the benchmark thermal conductivity. Inaccuracies in the third-order IFCs were simulated by either truncating digits or adding random numbers to the forces. The force errors were put in the irreducible set of IFCs before point/space group symmetry operations in order to make sure that the full set satisfies symmetry conditions. For truncation case, the force digits after a prescribed force accuracy value of each term were thrown. For example, when 0.1234567890123456 was truncated with a force accuracy of $10^{-3}$, the value would be 0.123. For random addition case, random numbers between the negative and positive force accuracy values were added to the force terms, which were generated with the rand function in standard C library. Five such random addition cases were conducted and the thermal conductivity was calculated for each of them. With the inaccuracies introduced, both original and modified third-order IFCs were used to calculate thermal conductivity. Modified third-order IFCs were produced by adding small compensation to each term using Lagrange multiplier method[12,54,58] in order to satisfy translational invariance conditions. In all the above-mentioned simulations, ShengBTE package[54] was used to calculate thermal conductivity with SMRTA. More simulation details that are related to different materials can be found in Sec. 3.

## 3. SIMULATION RESULTS

### 3.1 Silicon

Tersoff potential[59] was used to describe the Si-Si interactions in silicon. The face-centered cubic primitive unit cell and $3\times3\times3$ supercell were constructed and optimized successively. The lattice constant for equilibrium configuration with the lowest energy was found to be 5.4321 Å. Because a cutoff distance exists in the potential function, long-range IFCs are exactly zero. Accordingly, only second-nearest neighbors were considered in computing third-order IFCs.[60] Table 1 shows the maximum absolute values of translational invariance and rotational invariance tensors $\mathbf{T}^{(2)}$, $\mathbf{T}^{(3)}$, $\mathbf{I}^{(2)}-\mathbf{I}^{(2)\,\mathrm{T}}$, $\mathbf{I}^{(3)}-\mathbf{I}^{(3)\,\mathrm{T}}$(see the Appendix for the details). $\mathbf{T}^{(2)}$ and $\mathbf{I}^{(2)}-\mathbf{I}^{(2)\,\mathrm{T}}$ are the translational and rotational invariance tensors, respectively, for second-order IFCs. Meanwhile, $\mathbf{T}^{(3)}$ and $\mathbf{I}^{(3)}-\mathbf{I}^{(3)\,\mathrm{T}}$ are for third-order IFCs. “Original IFC3” in the first column refers to the case where nothing was added to the third-order IFCs and “Modified IFC3” refers to the case where small terms were added to make the third-order IFCs satisfy the translational invariance conditions. Modified IFC3 case is shown in the table merely for comparison purpose. Theoretically, these tensors should be equal to **0** if invariance conditions are strictly satisfied. However, the theoretical result is unattainable because numerical error always exists in computation. According to our experience, the results in Table 1 are much smaller than the commonly attainable values of first-principles calculations and we believe from these values that translational and rotational invariance conditions are well satisfied. By comparing the results between the two cases it can be found that both invariance conditions are better satisfied for modified IFC3. $36\times36\times36$ $q$ mesh was used to calculate the thermal conductivity at 300 K. Our results were about 319 W/mK for both cases and later we would take this value as the benchmark result for silicon. We noticed that previous calculation[60] based on the same method yielded a value about 15% higher than our result. Such a difference may arise from the better accuracy of our calculated IFCs and the denser $q$ mesh we have used. Furthermore, when comparing the results between the original IFC3 case and modified IFC3 case, we find that the difference is so small that it should be negligible, which might serve as the circumstantial evidence that our calculated thermal conductivity values are accurate.

Table 1 Translational, rotational invariance conditions and thermal conductivity of silicon from Tersoff potential (benchmark case)

| Case | $\left\|\mathrm{T}^{(2)}_{i\alpha\beta}\right\|_{\max}$ (eV/Å$^2$) | $\left\|\mathrm{T}^{(3)}_{ij\alpha\beta\gamma}\right\|_{\max}$ (eV/Å$^3$) | $\left\|\mathrm{I}^{(2)}_{i\alpha,\beta\eta}-\mathrm{I}^{(2)}_{i\alpha,\eta\beta}\right\|_{\max}$ (eV/Å) | $\left\|\mathrm{I}^{(3)}_{ij\alpha\beta,\gamma\eta}-\mathrm{I}^{(3)}_{ij\alpha\beta,\eta\gamma}\right\|_{\max}$ (eV/Å$^2$) | Thermal conductivity (W/mK) |
|---|---|---|---|---|---|
| Original IFC3 | $1.27\times10^{-12}$ | $3.37\times10^{-6}$ | $3.49\times10^{-7}$ | $6.48\times10^{-6}$ | 318.64 |
| Modified IFC3 | $1.27\times10^{-12}$ | $1.36\times10^{-9}$ | $3.49\times10^{-7}$ | $2.84\times10^{-6}$ | 318.78 |

Then we investigated the effect of the accuracy of third-order IFCs on the predicted thermal conductivity by changing the precision of forces, as shown in Fig. 1. The smaller value in the $x$ axis means better force accuracy. The horizontal solid gray line without symbols shows our benchmark result. Truncation cases are plotted as diamond and circular symbols connected with dashed line. The results from five random addition cases are plotted as floating bars in the figure. It can be seen from Fig. 1 that when the force accuracy value is on the order of or smaller than $10^{-10}$ eV/Å, all the calculated thermal conductivity values agree well with the benchmark result. The maximum error for these cases is less than 2% and therefore modified IFC3 is not quite necessary for getting an accurate result. When the force accuracy value is between $10^{-8}$ and $10^{-6}$ eV/Å, original IFC3 would yield 4.5%-20% error in thermal conductivity result but modified IFC3 can correct this error to less than 2%. For silicon, we take ±15% of the benchmark thermal conductivity as an acceptable range, and later we also use this criterion for graphene and silicene. For $10^{-5}$ eV/Å force accuracy case, modified IFC3 can reduce the error to this acceptable range for both truncation case and random addition cases. When the force accuracy is even worse, the absolute value of the error becomes larger than 15% and looks quite obvious in the figure. Therefore, we believe a reasonable thermal conductivity can be obtained for silicon from modified IFC3 with force accuracy at least on the order of $10^{-5}$ eV/Å.

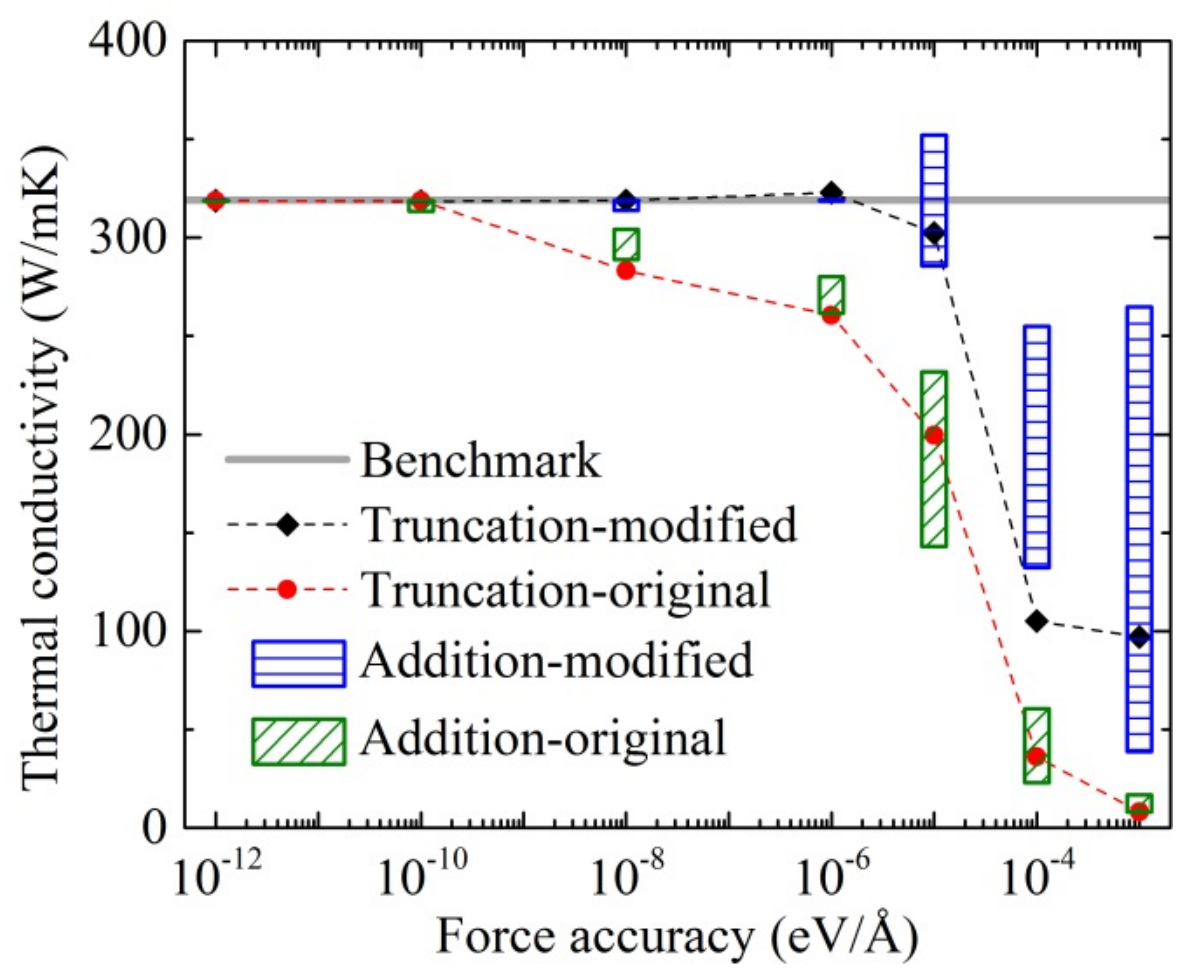

Fig. 1 Thermal conductivity of silicon from Tersoff potential vs. force accuracy

It can be seen that random addition cases show similar trend as truncation cases. The results from truncation cases are used for further discussions. Otherwise, five random addition cases need to be discussed for each force accuracy. When the force accuracy is on the order of $10^{-5}$ eV/Å, original IFC3 will yield translational invariance condition $\left|T^{(3)}_{ij\alpha\beta\gamma}\right|_{\max} = 0.468$ eV/Å$^3$ and modified IFC3 can reduce this value to $6.29\times10^{-10}$ eV/Å$^3$. Modification of third-order IFCs can make translational invariance conditions much better satisfied. Since it has been shown in Fig. 1 that modified IFC3, not original IFC3, gives reasonable thermal conductivity for this force accuracy, Grüneisen parameters from modified IFC3 cases are plotted in Fig. 2, together with those from second-order IFCs (labelled as IFC2). The $x$ axis indicates the reduced coordinates[61] of reciprocal lattice vectors. It can be seen from the figure that Grüneisen parameters calculated from IFC2 and benchmark IFC3 agree quite well, with the relative difference $e_{\gamma}$ =0.02% (see the Appendix for the definition of this relative difference). This can also serve as a circumstantial evidence that the third-order IFCs of the benchmark case are quite accurate. The modified IFC3 for $10^{-5}$ eV/Å accuracy still gives Grüneisen parameters that agree reasonably with IFC2, with a relative difference of 2.87%. It can be obviously seen from the figure that Grüneisen parameters for $10^{-4}$ eV/Å case are different from those for IFC2 case. Since $10^{-5}$ eV/Å force accuracy gives reasonable thermal conductivity, Grüneisen parameters may serve as one of the good criteria for the accuracy of modified third-order IFCs for silicon.

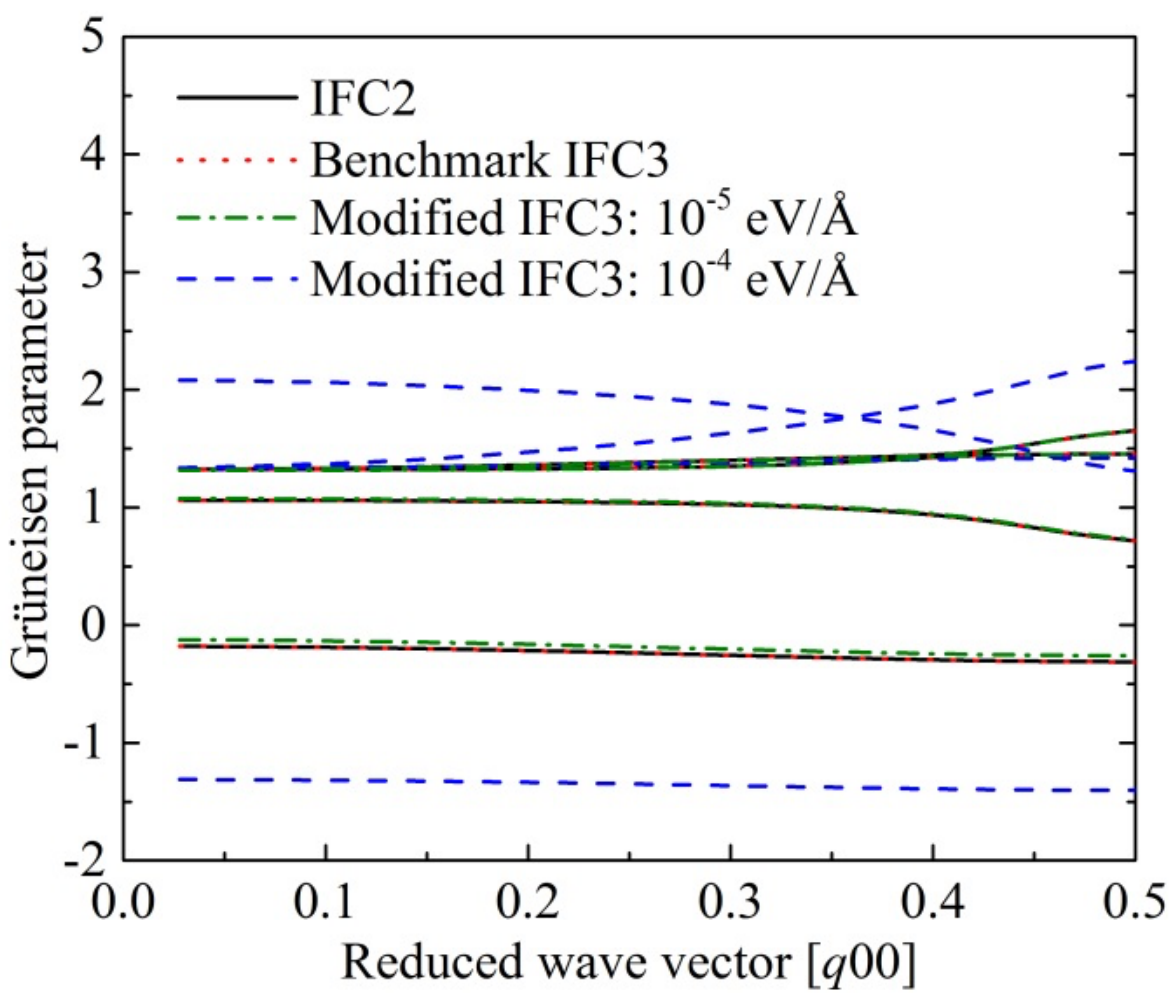

Fig. 2 Grüneisen parameters of silicon from Tersoff potential

### 3.2 Graphene

Graphene has attracted lots of scientific research interests since its discovery and has stimulated a growing research interest in 2D materials. The optimized Tersoff potential for graphene[62] was used to describe the C-C interactions. A hexagonal primitive unit cell structure was first constructed and optimized. Then a $5\times5\times1$ supercell was constructed and re-optimized. The lattice constant for the lowest energy state was finally found to be 2.492049 Å. For 2D materials, we used the original second-order IFCs to compute the phonon dispersion curve and check whether it would be consistent with the theoretical demonstration that a quadratic branch exists near Brillouin zone center.[48] The phonon frequencies for 1000 $q$ points from Γ to M were calculated and part of it was plotted in Fig. 3. Please note that the unit for $x$ axis is $2\pi/a$, where $a$ is the lattice constant. We can see from the figure that the dispersion curve for out-of-plane acoustic (ZA) mode looks like quadratic while the longitudinal and transverse acoustic (LA/TA) modes have linear dispersions. In order to further confirm the trend for ZA mode, the group velocities along Γ to M direction (derivatives of frequency) of ZA branch were calculated with finite difference method and were plotted as red circular dots in the inset of Fig. 3. The solid black line in the inset is a straight line crossing the origin point and the last point in the figure. It can be seen from the inset that all the dots for ZA group velocities sit on the straight line. Therefore, we believe the ZA phonon mode has a quadratic dispersion curve near Brillouin zone center, which agrees with the theoretical demonstration.[48]

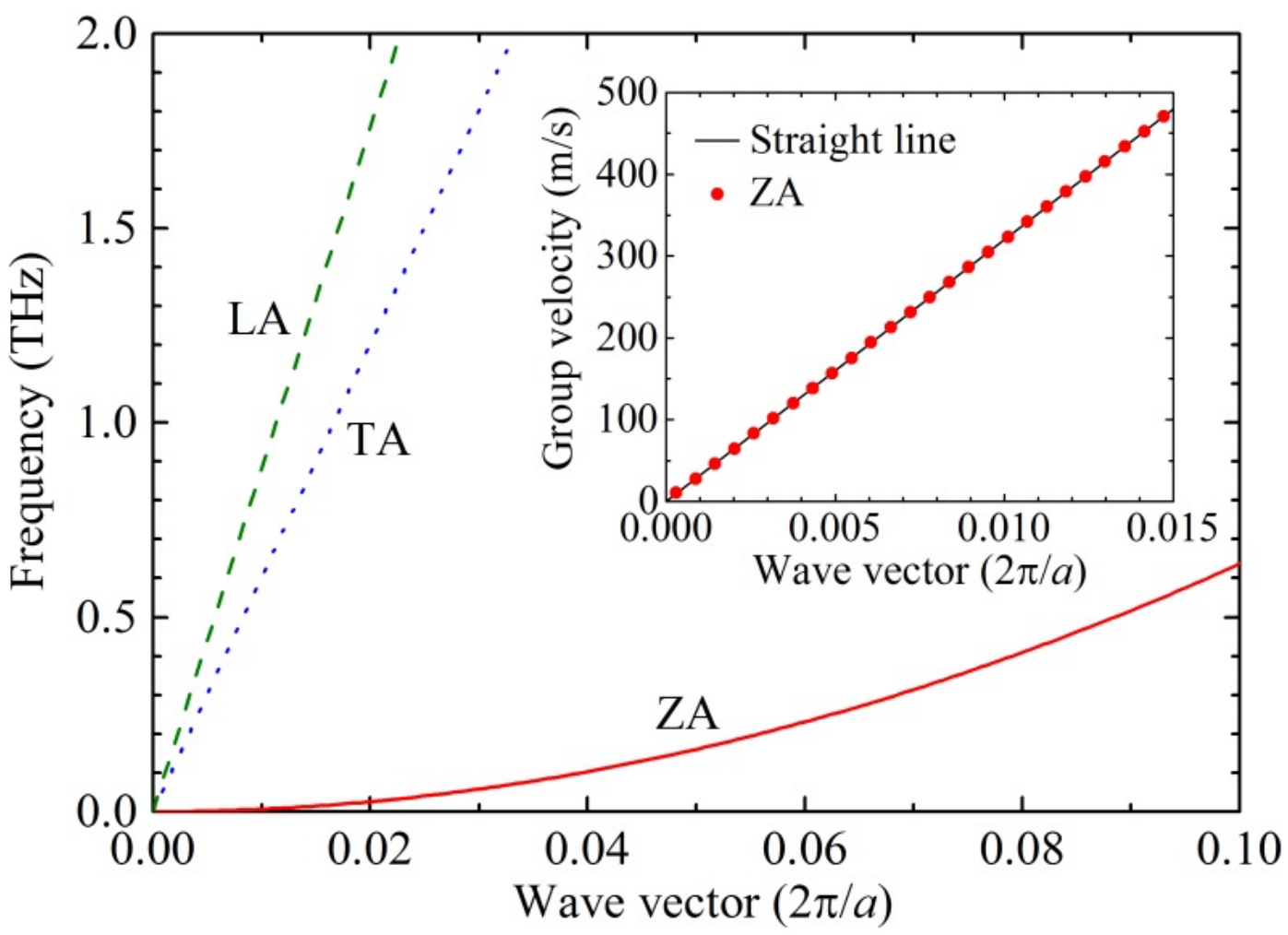

Fig. 3 Phonon dispersion curve of graphene calculated from optimized Tersoff potential for acoustic modes from Γ to M direction around Γ point. Inset: Group velocity of ZA mode along Γ to M direction around Γ point.

Regarding the calculation of third-order IFCs, second-nearest neighbors were considered. Table 2 shows the maximum absolute values of translational and rotational invariance tensors. The values of $\left|\mathrm{T}^{(3)}_{ij\alpha\beta\gamma}\right|_{\max}$ and $\left|\mathrm{I}^{(3)}_{ij\alpha\beta,\gamma\eta}-\mathrm{I}^{(3)}_{ij\alpha\beta,\eta\gamma}\right|_{\max}$ are larger than those of silicon, but we should notice that the C-C bonding in graphene is stronger than Si-Si bonding in silicon. These values are still very small either according to our experience or by comparing with the critical force accuracy case for silicon, *i.e.* $10^{-5}$ eV/Å case. $201\times201\times1$ $q$ mesh was used to calculate the thermal conductivity at 300 K. The results for original IFC3 and modified IFC3 cases are almost the same and we take the benchmark result as 633 W/mK. For graphene, using iterative method instead of SMRTA should give a much larger thermal conductivity[30] and it can be seen that the result from iterative method in Ref. [62] is much larger than our result here. Due to the reasons explained before, we just consistently use SMRTA method to do the accuracy analysis for silicon, graphene, and silicene.

Table 2 Translational, rotational invariance conditions and thermal conductivity of graphene from optimized Tersoff potential (benchmark case)

| Case | $\left\|\mathrm{T}^{(2)}_{i\alpha\beta}\right\|_{\max}$ (eV/Å²) | $\left\|\mathrm{T}^{(3)}_{ij\alpha\beta\gamma}\right\|_{\max}$ (eV/Å³) | $\left\|\mathrm{I}^{(2)}_{i\alpha,\beta\eta}-\mathrm{I}^{(2)}_{i\alpha,\eta\beta}\right\|_{\max}$ (eV/Å) | $\left\|\mathrm{I}^{(3)}_{ij\alpha\beta,\gamma\eta}-\mathrm{I}^{(3)}_{ij\alpha\beta,\eta\gamma}\right\|_{\max}$ (eV/Å²) | Thermal conductivity (W/mK) |
|---|---|---|---|---|---|
| Original IFC3 | $2.47\times10^{-12}$ | $3.87\times10^{-4}$ | $9.80\times10^{-6}$ | $4.82\times10^{-4}$ | 633.36 |
| Modified IFC3 | $2.47\times10^{-12}$ | $3.00\times10^{-9}$ | $9.80\times10^{-6}$ | $3.52\times10^{-5}$ | 633.49 |

Figure 4 shows our thermal conductivity result of graphene for different force accuracy. Please note that there is a break in the *y* axis because some results span a very large range. The horizontal gray line without symbols is the benchmark result. It can be seen that when the force accuracy value is on the order of or smaller than $10^{-5}$ eV/Å, all the calculated thermal conductivity values agree well with the benchmark result. Even the largest error is less than 5%. When the force accuracy is $10^{-4}$ eV/Å, the results from original IFC3 are within ±15% of the benchmark thermal conductivity, no matter for truncation cases or for random addition cases. In accordance with silicon, these results are taken as reasonable values. Anomalously, the results from modified IFC3 cases deviate more from the benchmark result than those from original IFC3. For truncation case, the error grows from 13% to 28%. For random addition case, modified IFC3 will yield a larger span of thermal conductivity from 555 to 812 W/mK than original IFC3 in the range of 551 to 671 W/mK. The results from modified IFC3 are out of the reasonable range we pre-defined. For the worst accuracy $10^{-3}$ eV/Å case studied here, it can be obviously seen from the figure that the discrepancy is quite large. Finally, we believe a reasonable thermal conductivity can be obtained for graphene from original IFC3 with force accuracy at least on the order of $10^{-4}$ eV/Å. This critical value of force accuracy for graphene is larger than that for silicon, which might arise from the stronger C-C bonding in graphene than the Si-Si bonding in silicon.

The accumulated thermal conductivity as a function of phonon frequency is helpful to understand the contributions of different phonons to the total thermal conductivity. We also show this plot for the benchmark case and $10^{-4}$ eV/Å truncation cases in Fig. 5. It can be seen that the three curves in the figure are smooth and show similar trend. It can be clearly seen that the inaccuracy of thermal conductivity is not due to a few phonon modes, but rather from the entire phonon spectrum.

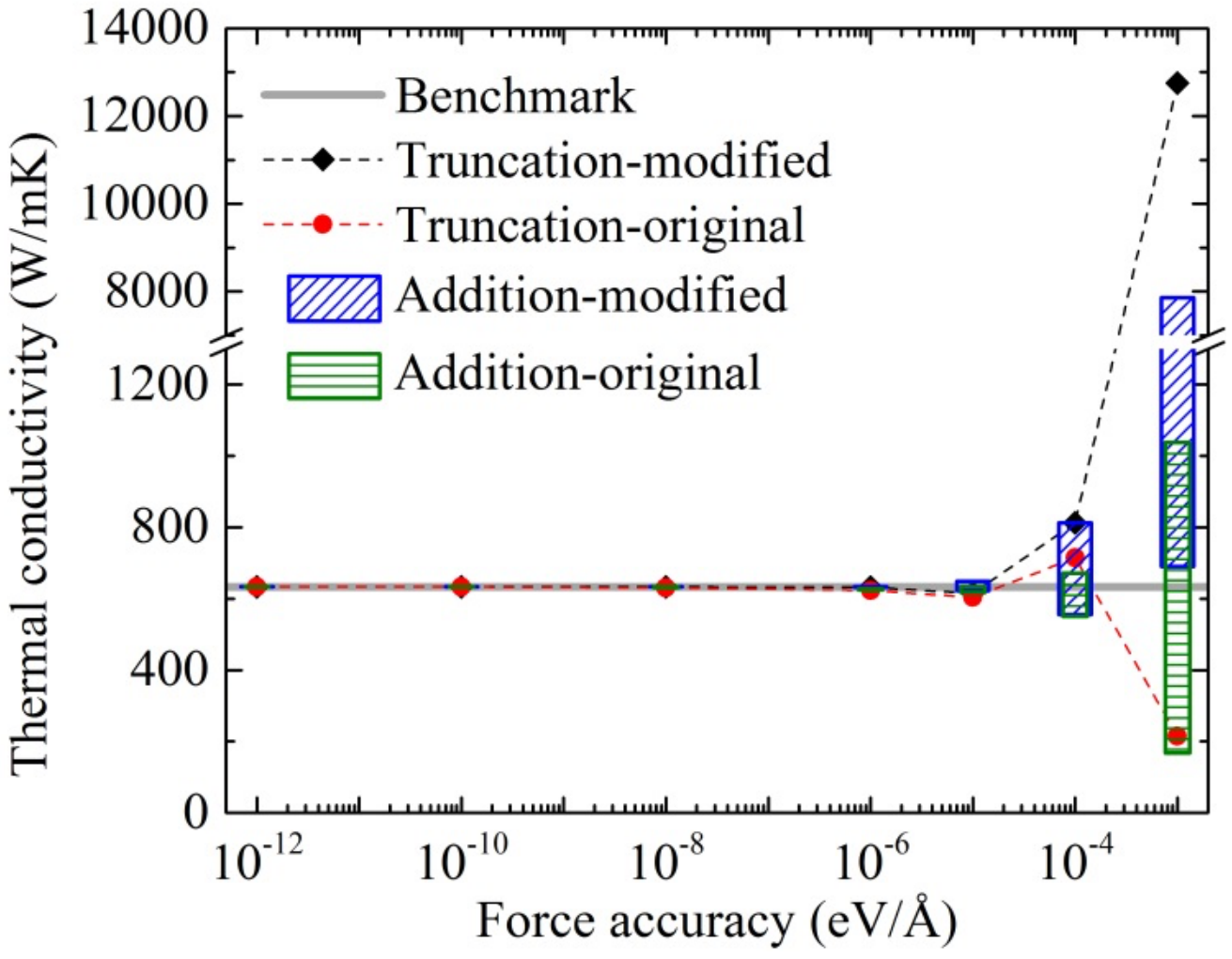

Fig. 4 Thermal conductivity of graphene from optimized Tersoff potential vs. force accuracy

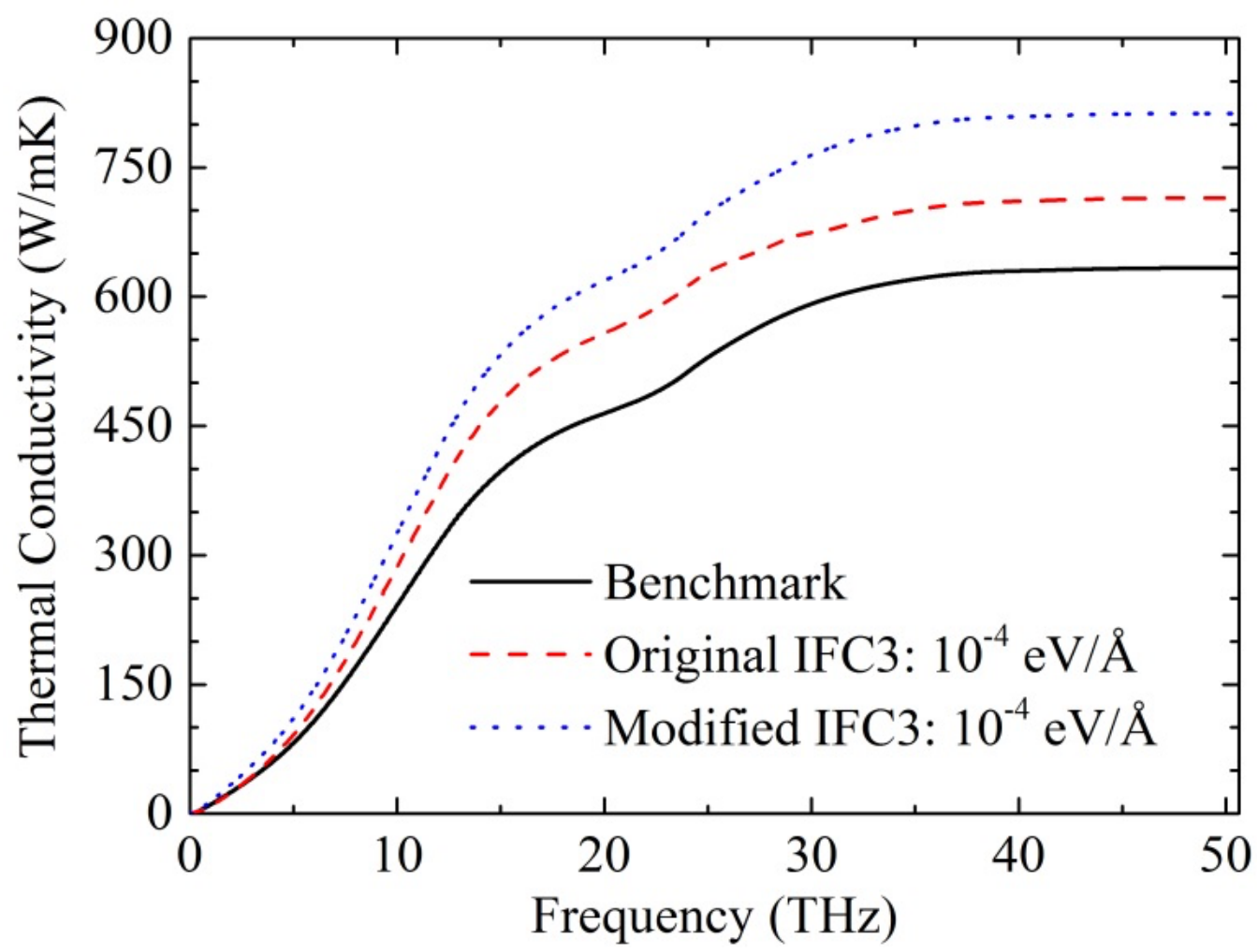

Fig. 5 Accumulated thermal conductivity of graphene as a function of phonon frequency

We still use the results from truncation cases for further discussions. From the Grüneisen parameters plotted in Fig. 6, it can be seen that the result from benchmark IFC3 agrees quite well with that from IFC2. The relative difference between them is only 2.01%. When the force accuracy is $10^{-4}$ eV/Å, original IFC3 will yield $\left|\mathrm{T}^{(3)}_{ij\alpha\beta\gamma}\right|_{\max} = 0.806\ \mathrm{eV/Å^3}$ and modified IFC3 can reduce this value to $1.96\times10^{-7}$ eV/Å$^3$. The translational invariance conditions are much better satisfied by modified third-order IFCs. Compared with original IFC3, Grüneisen parameters from modified IFC3 also show better agreement with those from IFC2. The relative difference of Grüneisen parameters is 22% between modified IFC3 and IFC2 and this is much larger than that between original IFC3 and IFC2. For graphene with $10^{-4}$ eV/Å accuracy, modification of third-order IFCs can reduce the error in translational invariance conditions and correct the trend of Grüneisen parameters, but does not yield a thermal conductivity that agrees better with benchmark result.

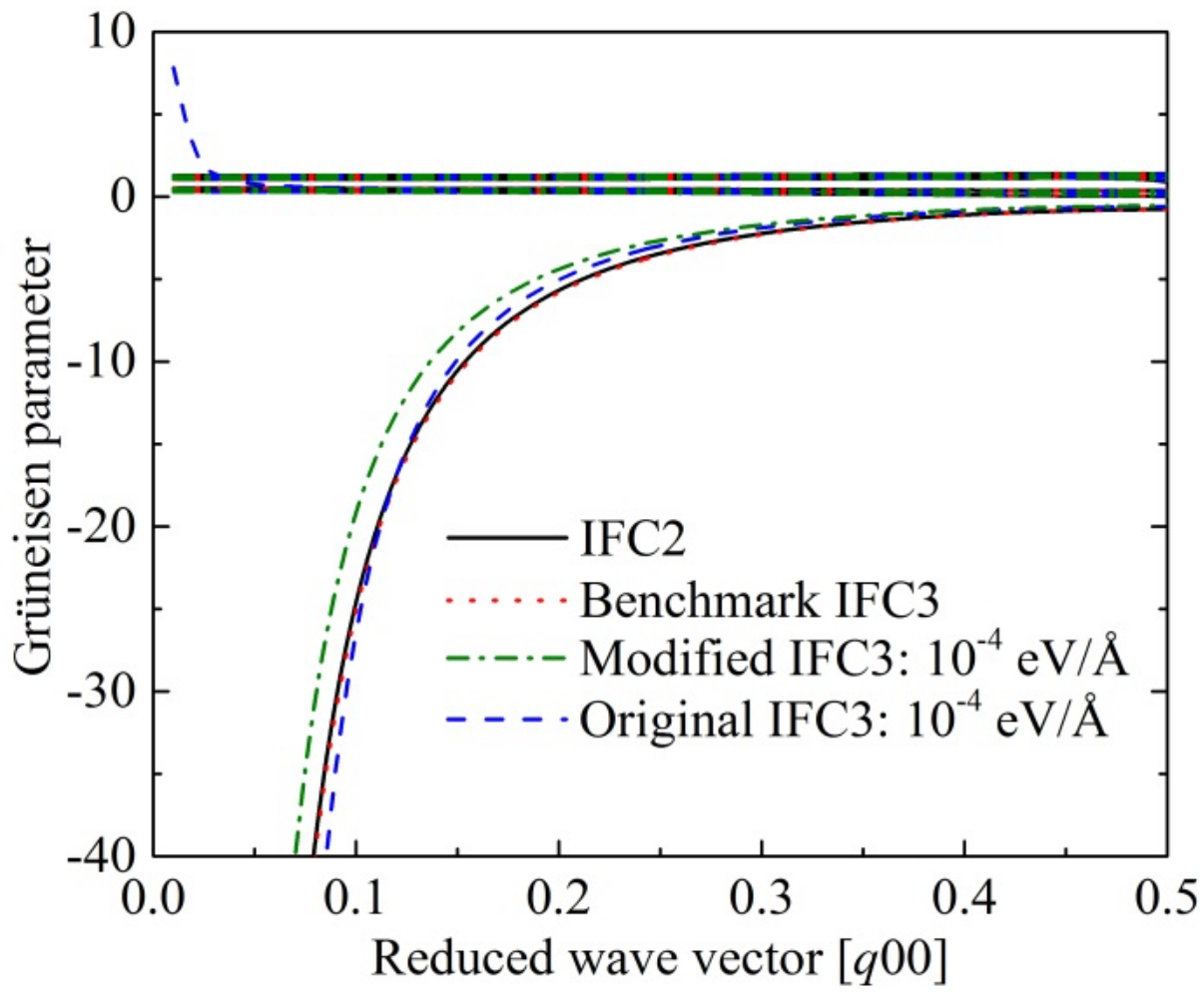

Fig. 6 Grüneisen parameters of graphene from optimized Tersoff potential

### 3.3 Silicene

Silicene is the counterpart material of graphene with a buckled structure, and therefore it was investigated with the same method as graphene. We used the Stillinger-Weber (SW) potential[63] to describe the Si-Si interactions and the optimized SW2 potential parameters for silicene from Ref. [64] was used. A buckled hexagonal primitive unit cell structure was first constructed for silicene and optimized. And then a $5\times5\times1$ supercell was constructed and re-optimized. The lattice constant for the optimized structure was 3.812425 Å and the buckling distance was 0.426948 Å. For silicene, we also calculated the phonon frequencies for 1000 $q$ points from Γ to M with original second-order IFCs. Our structure was well optimized and it could guarantee that no negative frequencies existed near Brillouin zone center even when we used so dense $q$ points, as shown in Fig. 7. The dispersion curves of acoustic modes for silicene show similar trends as those of graphene. It should be noted that the flexural acoustic (FA) mode in silicene is similar to the ZA mode in graphene but not purely out-of-plane.[32,35] It can be seen that LA/TA dispersion curves seem linear and FA dispersion curve looks like quadratic. We further plotted the group velocities of FA branch along Γ to M direction near Brillouin zone center as red circular dots in the inset of Fig. 7. The solid black line in the inset is also a straight line crossing the origin point and the last point in the figure. It can be seen from the inset that nearly all the red dots sit on the straight line, except that the first two points have a negligible deviation. These deviations are believed to arise from numerical uncertainty that brings very small residual strain. This result shows that the FA phonon for silicene also has a quadratic dispersion curve. Our previous results based on first-principles calculations[32,33,35] showed three linear dispersion curves for acoustic modes. Later Kuang *et al.*[34] showed that it would be quadratic from first-principles calculations. Actually, from our previous calculation based on SW potential, a quadratic branch was observed.[64]

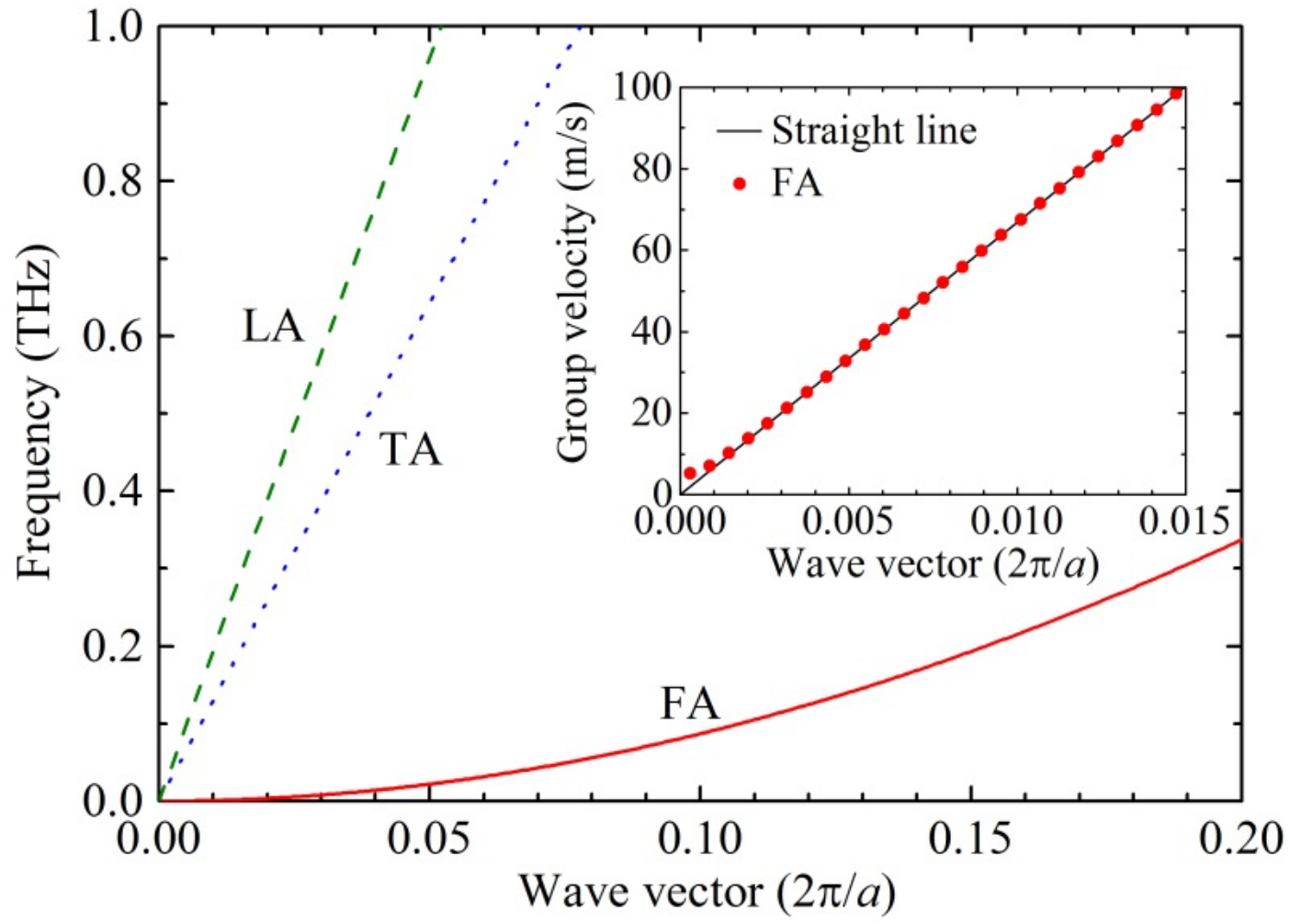

Fig. 7 Phonon dispersion curve of silicene calculated from optimized SW potential for acoustic modes from Γ to M direction around Γ point. Inset: Group velocity of FA mode along Γ to M direction around Γ point.

The second-nearest neighbors were considered in computing third-order IFCs. Table 3 shows the maximum absolute values of translational and rotational invariance tensors. These values are comparable to those of silicon and are also very small. Modified third-order IFCs better satisfy rotational and translational invariance conditions than original third-order IFCs. $201 \times 201 \times 1$ $q$ mesh was used to calculate the thermal conductivity at 300 K. The two results are exactly the same and we take the benchmark result as 11.94 W/mK. This result is higher than our previous calculation[64] because we used a much denser $q$ mesh compared with the previous $16 \times 16 \times 1$ one.

Table 3 Translational, rotational invariance conditions and thermal conductivity of silicene from optimized SW potential (benchmark case)

| Case | $\left\|T^{(2)}_{i\alpha\beta}\right\|_{\max}$ (eV/Å²) | $\left\|T^{(3)}_{ij\alpha\beta\gamma}\right\|_{\max}$ (eV/Å³) | $\left\|I^{(2)}_{i\alpha,\beta\eta} - I^{(2)}_{i\alpha,\eta\beta}\right\|_{\max}$ (eV/Å) | $\left\|I^{(3)}_{ij\alpha\beta,\gamma\eta} - I^{(3)}_{ij\alpha\beta,\eta\gamma}\right\|_{\max}$ (eV/Å²) | Thermal conductivity (W/mK) |
|---|---|---|---|---|---|
| Original IFC3 | $1.08\times10^{-14}$ | $2.39\times10^{-6}$ | $2.99\times10^{-6}$ | $4.56\times10^{-6}$ | 11.94 |
| Modified IFC3 | $1.08\times10^{-14}$ | $3.00\times10^{-10}$ | $2.99\times10^{-6}$ | $8.58\times10^{-7}$ | 11.94 |

Figure 8 shows our result of thermal conductivity for different force accuracy cases. The horizontal gray line without symbols also shows our benchmark result. It can be seen that when the force accuracy is on the order of or smaller than $10^{-8}$ eV/Å, all the calculated thermal conductivity values fully agree with the benchmark result. Even the maximum error is less than 0.2%. When the accuracy is $10^{-6}$ eV/Å, the error in thermal conductivity will be larger but modification of third-order IFCs can reduce it. For either truncation case or random addition case, thermal conductivity calculated from modified third-order IFCs is within ±15% of the benchmark result. When the force accuracy is even worse, the error for all the cases becomes very large and can be obviously seen in Fig. 8. Therefore, we believe a reasonable thermal conductivity can be obtained for silicene from modified IFC3 with force accuracy at least on the order of $10^{-6}$ eV/Å.

Truncation cases are used for further discussion about the accuracy of third-order IFCs. It is found that when the force accuracy is around $10^{-6}$ eV/Å for silicene, original third-order IFCs will yield translational invariance condition $\left|T^{(3)}_{ij\alpha\beta\gamma}\right|_{\max} = 0.0224$ eV/Å$^3$ and modified third-order IFCs can reduce this value to $5.36\times10^{-6}$ eV/Å$^3$. Grüneisen parameters are also plotted for IFC2 and modified IFC3 in Fig. 9. The Grüneisen parameters for all the cases shown in the figure seem to agree reasonably. The relative differences of Grüneisen parameters between IFC2 and IFC3 are less than 2% for the three different IFC3 sets. Considering that force accuracy on the order of $10^{-5}$ eV/Å does not yield a thermal conductivity result that agrees well with the benchmark result, we do not recommend using Grüneisen parameters as the criterion for the accuracy of modified third-order IFCs of silicene.

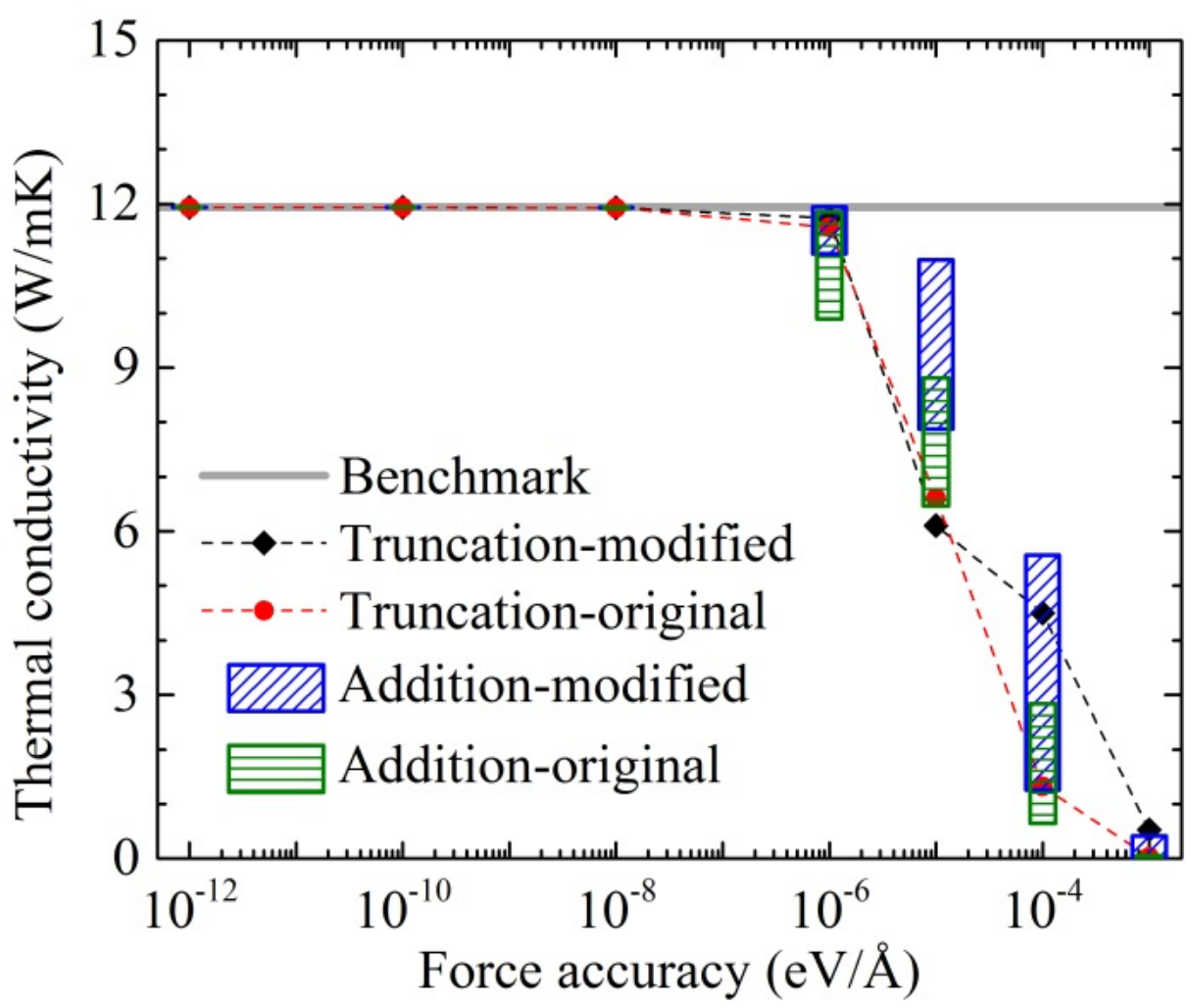

Fig. 8 Thermal conductivity of silicene from optimized SW potential vs. force accuracy

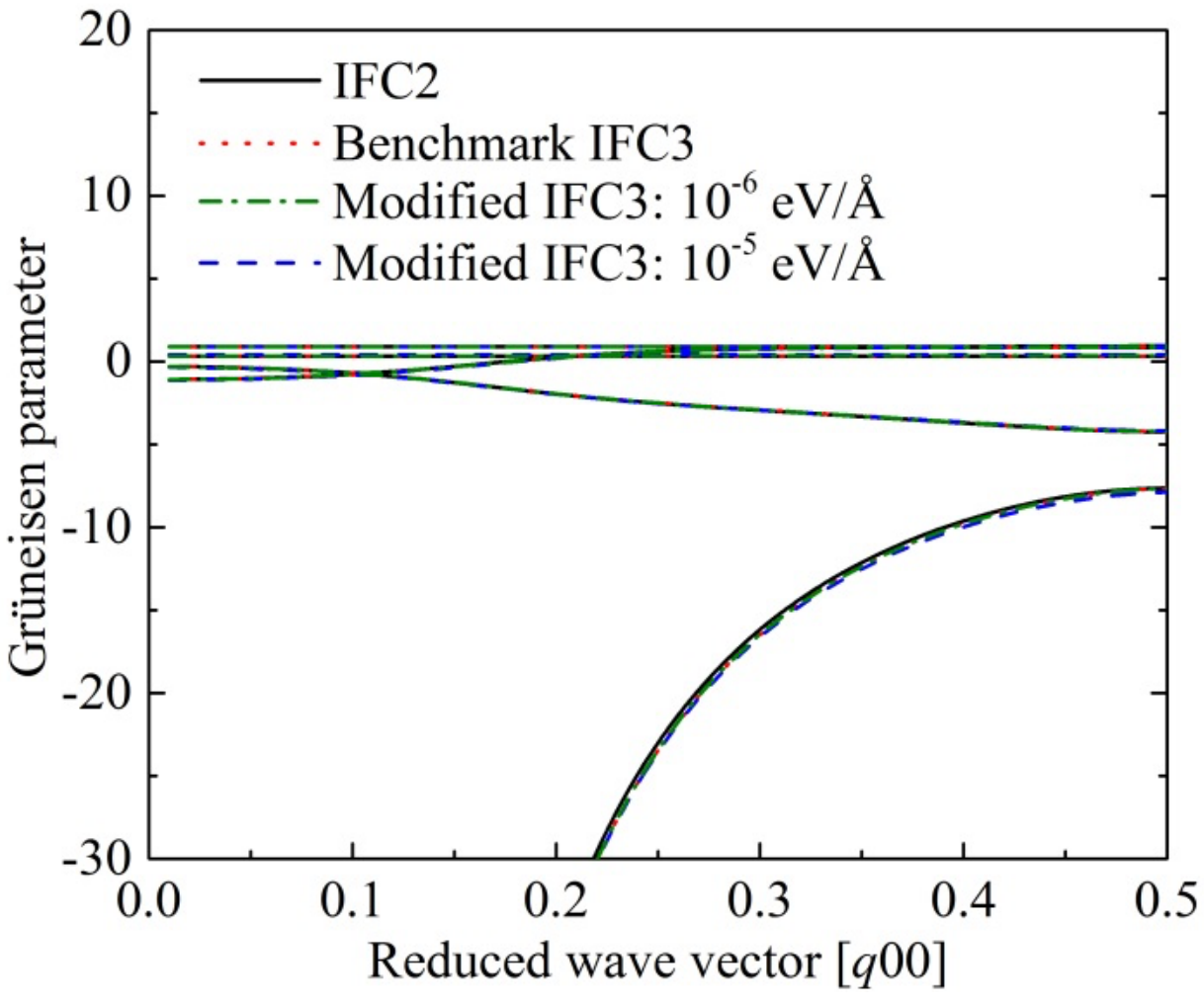

Fig. 9 Grüneisen parameters of silicene from optimized SW potential

## 4. DISCUSSION

Our benchmark thermal conductivity was calculated very carefully until the translational and rotational invariance conditions were well satisfied and Grüneisen parameters from second-order and third-order IFCs agreed quite well. For classical potential, a reasonable thermal conductivity can be obtained from modified IFC3 for silicon/silicene with force accuracy at least on the order of $10^{-5}/10^{-6}$ eV/Å. We can see that calculating the thermal conductivity of silicene requires higher accuracy of forces (and therefore third-order IFCs) than silicon, despite that the Si-Si interaction in silicon is stronger than that in silicene. For graphene, the required force accuracy is $10^{-4}$ eV/Å. Notably, when the force accuracy is $10^{-4}$ eV/Å for graphene, imposing translational invariance conditions to third-order IFCs does not give thermal conductivity that agrees better with benchmark value than using the original third-order IFCs. As shown in the Appendix, a finite displacement spacing $h$=0.01 Å was chosen in computing third-order IFCs, and the force error on the order of $10^{-4}/10^{-5}/10^{-6}$ eV/Å would introduce third-order IFC error approximately on the order of 1/0.1/0.01 eV/Å$^3$.

Regarding the first-principles calculations, it was noticed that large supercell and exceedingly high precision of forces were required to obtain accurate phonon properties.[49] For instance, discretization errors could be introduced from exchange correlation energy to forces and could be as large as 0.001 eV/Å per atom.[49] This would also cause the translational invariance conditions for forces to be destroyed[49] and therefore make the third-order IFCs inaccurate. The error in first-principles calculations may be affected by energy cutoff, $k$-point grid, reciprocal space projection technique, aliasing errors, discretization errors, and choice of control parameters.[49] When the force error introduced to classical-potential calculations lies in the typical value from first-principles calculations, our calculated thermal conductivity would be quite different from the benchmark value. This can explain the large differences of thermal conductivity values calculated from first-principles for 2D materials in literature. Especially, we have shown that very high accuracy of forces must be obtained in order to get accurate thermal conductivity values. Grüneisen parameters cannot be used as a necessary and sufficient criterion for the accuracy of third-order IFCs in the aspect of predicting thermal conductivity. One possible difference between classical potential and first-principles simulations is that the IFC range is limited by use of classical potentials. Due to the cutoff in classical potential, non-zero terms are within the second-nearest neighbors. In our calculations, enforcing translational invariance conditions only leads to changes in these non-zero terms[54]. While in first-principles calculations, more interacting neighbors are generally included and such changes can propagate into a larger number of IFCs.

## 5. SUMMARY

In summary, we used classical potential to investigate the effect of the accuracy of third-order IFCs on the predicted thermal conductivity based on SMRTA. It was found that when the force error lies in the typical value from first-principles calculations, the calculated thermal conductivity could be quite different from our benchmark result. Imposing translational invariance conditions was regarded as an effective approach to eliminate the inaccuracy in third-order IFCs. We found that such an effort could not always guarantee a better thermal conductivity result. Another criterion, Grüneisen parameters, has been used for the accuracy of third-order IFCs in the aspect of predicting thermal conductivity. We found that it could not be used as a necessary and sufficient criterion. Unfortunately we are not able to provide more instructive guidance on how to more accurately predict lattice thermal conductivity using the first-principles PBTE method. This is left for future investigations.

## ACKNOWLEDGEMENT

This work was supported by the National Natural Science Foundation of China (Grant No. 51676121) and the Materials Genome Initiative Center of Shanghai Jiao Tong University. Simulations were performed at High Performance Computing Center of Shanghai Jiao Tong University. We thank Fuman Xie (Shanghai Jiao Tong University) for her help in writing some of our code.

## APPENDIX

### A. Symmetry and invariance conditions for interatomic force constants

For a crystal with point/space group symmetries, the second-order and third-order IFCs must satisfy the following equations[50,51]

$$\Phi_{i'j'}^{\alpha'\beta'} = \sum_{\alpha\beta} \Phi_{ij}^{\alpha\beta} S^{\alpha'\alpha} S^{\beta'\beta} , \tag{3}$$

$$\Psi_{i'j'k'}^{\alpha'\beta'\gamma'} = \sum_{\alpha\beta\gamma} \Psi_{ijk}^{\alpha\beta\gamma} S^{\alpha'\alpha} S^{\beta'\beta} S^{\gamma'\gamma} , \tag{4}$$

where $\boldsymbol{S}$ is a point/space group symmetry operation matrix. $i'$, $j'$, and $k'$ are the atom indices corresponding to atom $i$, $j$, and $k$ after symmetry operation $\boldsymbol{S}$. These point and space group symmetries can be enforced by calculating an irreducible set of IFCs and then generating the other values according to Eq. (3) and (4), which can also reduce the computational cost. Under the fact that the potential energy would stay unchanged if the structure is translated as a whole under an arbitrary displacement, we can derive the translational invariance conditions (*i.e.* acoustic sum rules)[12,15,50,58,65,66] for second-order and third-order IFCs

$$\mathrm{T}_{i\alpha\beta}^{(2)} = \sum_{j} \Phi_{ij}^{\alpha\beta} = 0, \tag{5}$$

$$\mathrm{T}_{ij\alpha\beta\gamma}^{(3)} = \sum_{k} \Psi_{ijk}^{\alpha\beta\gamma} = 0 . \tag{6}$$

Similarly, rotational invariance conditions[50,51] can be derived under the fact that the potential energy would not change if the structure is rotated as a whole under an arbitrary angle, as shown below

$$\mathrm{I}_{i\alpha,\beta\eta}^{(2)} = \sum_{j} \Phi_{ij}^{\alpha\beta} r_j^{\eta} \text{ must be symmetric with respect to } \beta \text{ and } \eta , \tag{7}$$

$$\mathrm{I}_{ij\alpha\beta,\gamma\eta}^{(3)} = \sum_{k} \Psi_{ijk}^{\alpha\beta\gamma} r_k^{\eta} + \delta_{\alpha\eta} \Phi_{ij}^{\gamma\beta} + \delta_{\beta\eta} \Phi_{ij}^{\alpha\gamma} \text{ must be symmetric with respect to } \gamma \text{ and } \eta , \tag{8}$$

where $r_j^{\eta}$ is the Cartesian coordinate in $\eta$ direction of atom $j$ under equilibrium state. $\delta$ denotes the Kronecker delta function. $\mathbf{I}_{i\alpha}^{(2)} - \mathbf{I}_{i\alpha}^{(2)\,\mathrm{T}}$ and $\mathbf{I}_{ij\alpha\beta}^{(3)} - \mathbf{I}_{ij\alpha\beta}^{(3)\,\mathrm{T}}$ are implemented in my calculation and they should be equal to $\mathbf{0}$ if rotational invariance conditions are strictly satisfied. On condition that Eq.s (5)-(8) are not strictly satisfied, the IFCs are not physically accurate in the strictest sense. However, numerical error is unavoidable in computing IFCs and these analytical equations can never be strictly satisfied, *i.e.* satisfied with 0% error. Therefore the maximum absolute value of tensors $\mathbf{T}^{(2)}$, $\mathbf{T}^{(3)}$, $\mathbf{I}^{(2)}-\mathbf{I}^{(2)\,\mathrm{T}}$, $\mathbf{I}^{(3)}-\mathbf{I}^{(3)\,\mathrm{T}}$ are used to examine to what extent translational and rotational invariance conditions are satisfied.

### B. Grüneisen parameters

Grüneisen parameters can be calculated from second-order IFCs with the following equation[60]

$$\gamma_{\lambda}^{(2)}=-\frac{V}{\omega_{\lambda}}\frac{\Delta\omega_{\lambda}}{\Delta V}, \tag{9}$$

where $\omega_{\lambda}$ is the phonon frequency and can be obtained from second-order IFCs. $V$ is the crystal volume. Alternatively, Grüneisen parameters can also be calculated from third-order IFCs[60]

$$\gamma_{\lambda}^{(3)}=-\frac{1}{6\omega_{\lambda}^{2}}\sum_{ijk,\alpha\beta\gamma}\Psi_{ijk}^{\alpha\beta\gamma}\frac{\varepsilon_{i\lambda}^{\alpha\,*}\varepsilon_{j\lambda}^{\beta}}{\sqrt{m_i m_j}}\exp\left(\boldsymbol{iq}\cdot\mathbf{R}_j\right)r_k^{\gamma}, \tag{10}$$

where $\boldsymbol{\varepsilon}_{i\lambda}$ is the phonon eigenvector and the * denotes the complex conjugate. $m_i$ is the atomic mass of the $i$-th atom. $\mathbf{R}_j$ is the unit cell vector of the unit cell where the $j$-th atom locates. It should be noted that the $i$ in the exponential parenthesis is the imaginary unit instead of an atom index. $\gamma_{\lambda}^{(2)}$ and $\gamma_{\lambda}^{(3)}$ were calculated with Phonopy package[67] and ShengBTE package, respectively. The relative difference between the Grüneisen parameters calculated from second-order IFCs $\gamma_{\lambda}^{(2)}$ and those calculated from third-order IFCs $\gamma_{\lambda}^{(3)}$ is defined as $e_{\gamma}=\sqrt{\frac{\sum_{\lambda}\left|\gamma_{\lambda}^{(2)}-\gamma_{\lambda}^{(3)}\right|^{2}}{\sum_{\lambda}\left|\gamma_{\lambda}^{(2)}\right|^{2}}}$.

### C. Numerical error in finite difference method

In general, IFCs are computed with finite difference method. For classical potential, the error mainly comes from rounding error and truncation error. Rounding error results from the fact that floating-point numbers are represented in a computer by a finite number of digits of precision.[68] Using double-precision floating-point numbers in C/C++ programming would yield negligible rounding error. Truncation error comes from the finite difference method. The second-order and third-order IFCs can be calculated with[69]

(a) central difference method

$$\Phi_{ij}^{\alpha\beta} = -\frac{1}{h}\sum_{m=-1}^{1} C_m F_i^{\alpha}\left(r_j^{\beta} + mh\right) + O\left(h^2\right), \tag{11}$$

$$\Psi_{ijk}^{\alpha\beta\gamma} = -\frac{1}{h^2}\sum_{m=-1}^{1}\sum_{n=-1}^{1} C_m C_n F_i^{\alpha}\left(r_j^{\beta} + mh, r_k^{\gamma} + nh\right) + O\left(h^2\right), \tag{12}$$

with coefficients $C_{-1} = -\frac{1}{2}, C_0 = 0, C_1 = \frac{1}{2}$. $h$ is the finite displacement spacing. As defined before, $\boldsymbol{F}_i$ is the force acting on atom $i$. $O\left(h^2\right)$ denotes a negligible term of the order $h^2$.

(b) finite difference method with fourth-order accuracy $O\left(h^4\right)$

$$\Phi_{ij}^{\alpha\beta} = -\frac{1}{h}\sum_{m=-2}^{2} C_m F_i^{\alpha}\left(r_j^{\beta} + mh\right) + O\left(h^4\right), \tag{13}$$

$$\Psi_{ijk}^{\alpha\beta\gamma} = -\frac{1}{h^2}\sum_{m=-2}^{2}\sum_{n=-2}^{2} C_m C_n F_i^{\alpha}\left(r_j^{\beta} + mh, r_k^{\gamma} + nh\right) + O\left(h^4\right), \tag{14}$$

with coefficients $C_{-2} = \frac{1}{12}, C_{-1} = -\frac{8}{12}, C_0 = 0, C_1 = \frac{8}{12}, C_2 = -\frac{1}{12}$. $O\left(h^4\right)$ denotes a negligible term of the order $h^4$. Using fourth-order accuracy finite difference method instead of central difference method can reduce the truncation error but requires calculating more cases for forces and thus would lead to more computational cost. In our calculations, we chose $h$=0.01 Å and believed it to be a reasonable choice based on our test. With errors introduced to the forces, they will propagate to the third-order IFCs. When the maximum error introduced to the force term is $\Delta F$, the maximum third-order IFC error will be $\Delta F/h^2$. This can be understood because (i) if $\Delta F$ is large (e.g. larger than $10^{-12}$ eV/Å), $O\left(h^4\right)$ will be much smaller than $\Delta F/h^2$ and thus is negligible (ii) if $\Delta F$ is small (e.g. smaller than $10^{-12}$ eV/Å), the coefficients of $O\left(h^4\right)$ will not change much and they will be cancelled out if we use the inaccurate third-order IFC to subtract the benchmark accurate value.